*De novo oscillationum genere*

On a new class of oscillations

Author

Georg Wolfgang Krafft

*Commentarii academiae scientiarum Petropolitanae, Tomus X*, 1738, pp. 200-206

*Emendanda. Tab. XVI fig. 1. et fig. 2.*

Translated by

Sylvio R. Bistafa[a]

January 2021

Foreword

This publication was motivated by Krafft's accidental observation of a suspended clock setting itself in constant motion as a pendulum. His analysis of the phenomena led him to conclude that the vibration of the clock was solely due to unbalances in the clock's balance wheel. Next, he conceives a 'little machine' as he called it, in which a straight bar loaded by small weights at its extremities is free to oscillate about the center of gravity of a regular suspended clock. He then investigates different oscillating conditions, by calculating moments with different weights and lengths of the bar arms, to find conditions to attain oscillating excursions in a right angle, excursions with maximum amplitude, and showing that the vibrations of the whole clock are rendered more sensible, the shorter is the height of the suspension.

§1.

Although, different cases in which the oscillatory motion can be produced in bodies had been reported to this day, such as: of a body hanged up by string, of a body hanging down from some of its point, of the water in a tube driven by a communicating reciprocal motion, of a tendon or a string strongly extended and pinched by a spike; and many others: yet, another type of oscillation that deserves attention, which is due to the internal structure of the body, which nobody has examined; recommends itself not only by its elegance, but also by the perseverance of its constant motion.

§2. Having been accidentally led into the investigation of a new oscillatory motion, which happened in the portable clock, hanging up freely from a hook, such that its back surface does not come in contact with the wall, I had observed harmony in these oscillatory motions, which describe arcs of quite sensible magnitudes, that for the first time grabbed my admiration in a considerable degree, and, afterwards, gave me the opportunity to think over them, and once having been easily induced into the cause of this phenomenon, I had, simultaneously, extended a little broadly, particularly on the origins of these oscillations.

§3. Since any weight hanging down vertically cannot stay at rest upon an action, unless an opposing reacting force is readily applied: it is evident that if a body suspended in its center of gravity were to stay at rest, it is necessary that an insurmountable obstacle be placed at the said point, and in the same

---

[a] sbistafa@usp.br

vertical straight line in which the body is suspended, and to be free of oscillations , from which it may behave just like all simple pendulums continuously striving, as soon as they have touch that [obstacle], they will cease to oscillate, and will be brought back to rest.

§4. Accordingly, in order to obtain an explanation to the phenomenon, be set $ABKL$ as a section of a clock (*Tab. XVI fig. 1.*), which will be conceived as being circular, with its center of gravity placed in the center $C$ of the circle; be placed to its side a balance wheel $GBF$, which is constructed as usual, and consisting of a hardened ring, with three radius $MF, MG, MB$, irradiating from its center $M$, which divide the circumference in three equal arcs. In the extremities $F, G, and\ B$, of these radiuses, are usually attached devices of some kind of metal to enhance the vibration of the balance wheel. If, however, it is constructed such that the ring has the same width and thickness everywhere, and, moreover, if the small weights are constructed such that they have precisely the same weight: then the center of gravity of this balance wheel will be at the center $M$ of the circle, and it will remain at the same fixed point $M$ in case the balance wheel vibrates or not. But, if it happens that either the ring is not uniform or the weights are not equal, or both: the center of gravity of the balance wheel will be located at a point out of the center, which is set at $I$. Then, having been joined the center of gravity of the clock without the balance wheel $C$, and that of the balance wheel $M$, by a straight segment $CI$, then, the common center of gravity will be at some intermediate point $E$; which, if it is in the same vertical $DE$ with the suspension $D$: then the whole system will be at rest, with the balance wheel remaining at this location. Now, the balance wheel is set into motion, with the radiuses of first vibration reaching the locations $fM, gM, hM$, so that the location of the center of gravity $I$ ascends through the small arc $Ii$, described by the radius $MI$ with center at $M$; furthermore, the common center of gravity $E$ will reach another location $e$, making another vertical line $eD$: the clock will no longer stay at rest at that position, and an oscillation will take place, which will endure until the point $e$ reaches the vertical $DE$. Therefore, it is evident that the oscillatory motion originated in such a clock, is solely due either to the differences in thickness of the balance wheel, or the inequalities of the small weights attached at $F, G, and\ H$.

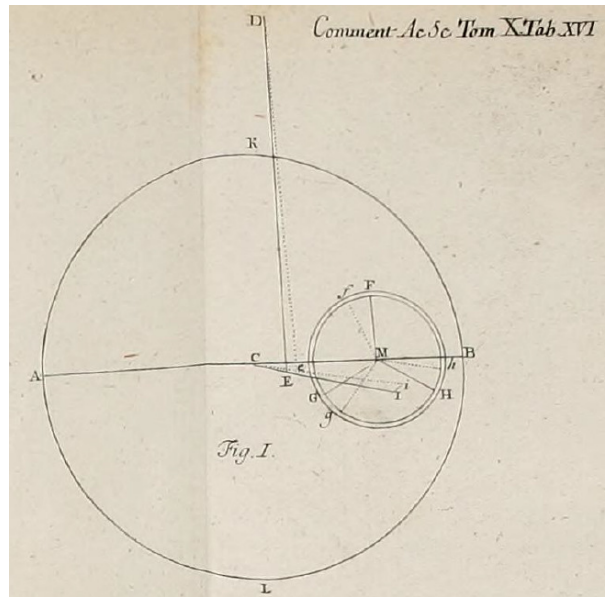

§5. We now conceive a new little machine (*Tab. XVI fig. 2.*), which resembles such a clock, from a circular thin sheet of metal $RLMN$, in which a bar of unequal arms and lacking gravity vibrates about its center $C$, and whose extremities are loaded with small weights $A$ and $B$; however, the diverse vibrations generated by the spring and the small wheels, no matter how they are arranged, are such that the center of gravity of the thin sheet together with the small wheels and the spring can be considered as being in the actual center $C$, and although it may be elsewhere, for the sake of simplicity, I assume that it is not much far apart. Now set $AC = a, CB = b$, weight of the thin sheet and of the small wheels $= C$, and weight attached to both sides of the arm $= A$ and $= B$; the common center of gravity between $C$ and $A$ is sought; this will be found out by analogy $A + C: A = a: EC$, whence $EC = \frac{Aa}{A+C}$ and also, by a similar analogy, the common center of gravity between $E$ and $B$ will be obtained by putting $A + B + C: B = EB: ED$, whence $ED = \frac{B \cdot EB}{A+B+C}$, or, since $EB = AB - AE = a + b - \frac{Ca}{A+C} = \frac{Aa+Ab+Cb}{A+C}$, then $ED = \frac{ABa+ABb+BCbEB}{(A+C)(A+B+C)}$, consequently, furthermore, $CD = ED - EC = \frac{(A+C)(Bb-Aa)}{(A+C)(A+B+C)} = \frac{Bb-Aa}{A+B+C}$, from which, the point $D$ is obtained, and here lies the location of the center of gravity of the whole little machine.

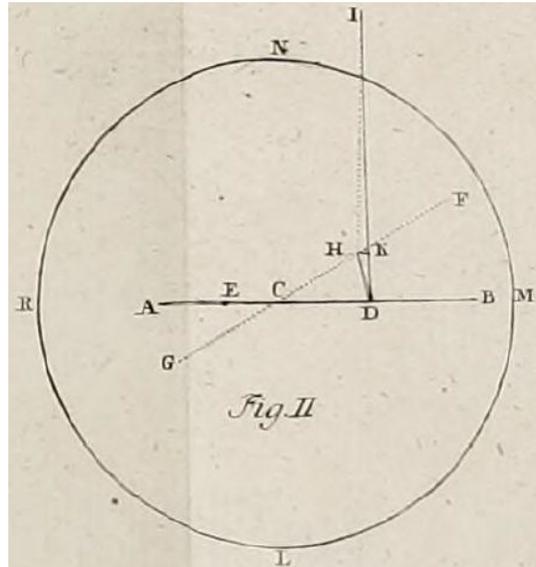

*Fig. II*

§6. This little machine is now suspended at point $D$ through the suspension point $I$: it will remain in equilibrium, with no oscillations. But if thereafter a vibratory motion initiates at the bar $AB$, induced by the attached spring and the small wheels, then, the first vibration will bring the bar to the location $GF$, in which the common center of gravity $D$ rises to the position $H$, completing afterward a circular arc $DH$ with center at $C$; and since the point $H$ does not belong to the same vertical $DI$: an oscillatory motion will follow towards $M$, and as long as the point $H$ remains at the said vertical, several oscillations will be accomplished, until the machine reposes again, unless the vibrations of the bar endure.

§7. Moreover, in order to obtain the angle $HID$, may be guided the chord $HD$, and the segment $HK$, which is parallel to $AB$; and the angles $H$ and $D$ will be equal; hence, calling the sine of $FCB$ by $m$, the cosine by $n$, $ID$ by $e$, then $[angle]\ CDH = [angle]\ CHD$, the sine is $\frac{\sqrt{1+n}}{\sqrt{2}}$, the cosine is $\frac{\sqrt{1-n}}{\sqrt{2}}$; therefore,

from the proportions, $\frac{\sin CHD}{CD} = \frac{\frac{\sqrt{1+n}}{\sqrt{2}}}{\frac{Bb-Aa}{A+B+C}} = \frac{\sin CHD(m)}{DH}$, giving the chord $DH = \frac{(Bb-Aa)m\sqrt{2}}{(A+B+C)\sqrt{1+n}}$, henceforth, since the angle $HDI$ is the complement of the angle $CDH$, its sinus is $= \frac{\sqrt{1-n}}{\sqrt{2}}$; whereby in the triangle $HKD$ we have that $\frac{1}{DH} = \frac{\sin HDI}{HK} \left[ because \sin HDI = \frac{HK}{HD} \right]$, then, $HK = \frac{DH\sqrt{1-n}}{\sqrt{2}}$; likewise, $\frac{1}{DH} = \frac{\sin DHK}{HK} \left[ because \sin DHK = \frac{DK}{HD} \right]$, then, $DK = \frac{DH\sqrt{1+n}}{\sqrt{2}}$; whence we have that $IK = ID - DK = e - \frac{DH\sqrt{1+n}}{\sqrt{2}}$, and, consequently, the tangent of the angle $HID = \frac{HK}{IK} = \frac{DH\sqrt{1-n}}{e\sqrt{2}-DH\sqrt{1+n}}$, or, by substituting the value of $DH$ found above, results in $\tan HID = \frac{(Bb-Aa)m}{et(A+B+C)-mt(Bb-Aa)}$, knowing that $\frac{\sqrt{1+n}}{\sqrt{1-n}} = t$, which is the tangent of the angle $CDH$.

§8. In case this tangent vanishes, then, the little machine loses its oscillatory motion for $Bb = Aa$, or, indeed, $\frac{A}{B} = \frac{b}{a}$, that is, if the weights $A$ and $B$ attached to the bar are as the reciprocal ratio of their distances to the fulcrum $C$. Thus, in this case, the common center of gravity falls on an immutable position $C$, which is not changed by the vibrations of the bar.

§9. If one looks for the value of $m$ as due to the angle $HID$ for which the vibrations make a maximum excursion, this will be found by putting $Bb - Aa = \alpha$, $A + B + C = \beta$, giving the equation (A)[b] as $\beta etdm - \beta emdt + \alpha m^2 dt = 0$; and since $t = \frac{\sqrt{1+n}}{\sqrt{1-n}}$, then, $dt = \frac{-mdm}{mn(1-n)}$, which when substituted into equation (A), and after dividing the resulting expression by $dm$, gives $\beta et + \frac{\beta em}{n(1-n)} - \frac{\alpha m^2}{n(1-n)} = 0$, and by substituting the value of $t$ into this expression, and after multiplying the resulting expression by $\sqrt{(1-n)}$, and dividing it by $m$, gives $\beta en + \beta e - \alpha m = 0$, or else, $\beta e\sqrt{(1-m^2)} + \beta e - \alpha m = 0$, finally resulting in $m = \frac{2\alpha\beta e}{\alpha^2+\beta^2 e^2} = \frac{2e(Bb-Aa)(A+B+C)}{(Bb-Aa)^2+e^2(A+B+C)^2}$. Therefore, the angle for which the vibrations make a maximum excursion, and to detect the largest vibrating amplitude of the arm $AB$, its sine should have the alleged value $m$.

§10. For the oscillations to make an excursion in a right angle, it is required that the tangent of $HID$ be infinite, or $et(A+B+C) - mt(Bb-Aa) = 0$, whence giving $e = \frac{m(Bb-Aa)}{A+B+C}$.

§11. In the traditional portable clocks, such as those made before Huygens's invention, in which the balance bar was loaded on both sides with equal small weights, and with arms of equal length, which corresponds to put $a = b$ in the expression just found for the angle $HID$, whence, the tangent of the angle $HID$ will be $\tan HID = \frac{(B-A)am}{et(A+B+C)-amt(B-A)}$. Since, indeed, $A + B$ is almost zero when compared to $C$, and also, when compared to $C$, $B - A$ is a negligible quantity: the tangent of the said angle $HID$ will be given by $\tan HID = \frac{(B-A)am}{Cet}$, therefore, since in any given clock the quantities $B, A, C, a, m, t$ are all constants, then, only the height of the suspension $e$ is possible to vary; then becomes clear that the tangent of the angle covered by the oscillations in such a clock is inversely proportional to the height of

---

[b] TN: A classical search for maximum, by equating the differential to zero.

the suspension, or, for that reason, the vibrations are rendered more sensible, the shorter is the height of the suspension.

§12. Since the weight $C$ can be freely increased or reduced, of course, a new weight can be added to point $C$, or a certain portion similar to the one added can be eliminated from the circular thin sheet, such that the center of gravity is always at point $C$, let us set $C = B(b-1) - A(a+1)$; which when substituted into the expression for the tangent of $HID$ found above, will give this expression $\tan HID = \frac{(Bb-Aa)m}{et(A+B+C)-mt(Bb-Aa)} = \frac{m}{et-mt} = \frac{m}{(e-m)t}$; in this case, the angle of excursion $DIH$ turns out independent of the small weights $A, B, C$; yet, the little machine cannot be deprived from oscillating, unless $m = 0$, that is, unless the balance wheel stays at rest. In this same case, for the angle $HID$ to become maximum, it is necessary that $m = \frac{2\alpha\beta e}{\alpha^2+\beta^2 e^2} = \frac{2e(Bb-Aa)(A+B+C)}{(Bb-Aa)^2+e^2(A+B+C)^2} = \frac{2e}{e^2+1}$; and for this same angle to become a right angle, it is necessary that $e = m$.

§13. If the bar of unequal arms $AB$ is replaced by the homogeneous bar $CB$, then $A = 0$, and $a = 0$, whence, the tangent of the angle $HID$ will be $\tan HID = \frac{Bbm}{et(B+C)-mtBb}$, then, it becomes clear that in this case, the little machine cannot be freed from oscillating, unless, again, $m = 0$, that is, the balance wheel stays at rest.

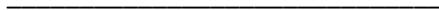